\documentclass[11pt]{journal}
\usepackage{fullpage}
\usepackage[english]{babel}
\usepackage[utf8x]{inputenc}
\usepackage{cprotect}
\usepackage [autostyle, english = american]{csquotes}
\usepackage{dirtytalk}
\usepackage[letterpaper, portrait, margin=1in]{geometry}

\usepackage{amsmath}
\usepackage{graphicx}
\usepackage[colorlinks=true, allcolors=blue]{hyperref}
\usepackage{amstext} 
\usepackage{array}
\usepackage{gensymb}
\usepackage{url}
\usepackage{graphicx}
\title{Improving Medical Systems in the United States using Knowledge-Based  Systems}
\author{\bf{Seongwoo Choi} (Class of 2017) \\ University of California, Santa Cruz \\ Computer Science BS}
\date{}

\begin{document}
\maketitle
\section*{Abstract}
America has one of the best medical systems in the world. The medical treatment care options offered by the medical system make it sophisticated. However, many American patients are not receiving health care on a regular basis, and at the same time, they cannot afford it. Also, the current medical system has many flaws such as high medical treatment costs and lack of doctors to accommodate many patients. This paper presents the principles of medical artificial intelligence called the “knowledge-based system”. Doctors can remotely check and monitor their patients' health data, medical history, how and what medical tests were done, and the lab results. The patients have access to detailed health information online and do not need to make an appointment with doctors to check their health on a daily basis.   One doctor can check many patients simultaneously online (when medical centers are understaffed) and do not need to spend a lot of time with patients. Thus, doctors save more money for patients, because patients will no longer be transporting to medical centers to receive routine health check-ups. Patients do not need to overpay for their insurance because they will have access to the “knowledge-based system,” and the system will save the patient’s money to have their health checked and reduce the number of unnecessary medical exams.  This paper undertakes a brief overview of research work done in a knowledge-based system / rule-based expert systems in the field of medical practices.
  
\section{Introduction}
	The interest of health in America continues to be a major issue as more people are becoming more interested in their health care plans and they are demanding more from the United States government to come up with a better and more affordable health care plan so that people can get treatment for their various illnesses and diseases. Although many people want affordable and effective medical insurance plans, it is hard to fix this epidemic due to the lack of health care infrastructures and appropriate treatments available. These two things tend to exacerbate each other. For instance, if a person drinks contaminated water, he or she is more likely to contract an illness or disease. On the other hand, people who do not have access to health care are not as informed about the harms of contaminated water.
     It is of the utmost importance that every individual has access to the affordable and easy solution to get health care from doctors. Maintaining good health is the basis for human and social development. With good health, people are more productive, happy, and enjoy a better quality of life. It is during times like this that people can function to their fullest potential. However, illnesses can come suddenly. It is during this time that people need the most medical treatment from a medical center and their health care plans so that people can get medical treatment at as low a price as possible. However, universal access to health care is prevented by the high costs associated with medical care. Most people, even in the United States, are unable to pay for their medical treatment fully.  Rather, they rely primarily on the government or some insurance contract to provide their treatments. Even this, however, cannot solve the problem of the high costs associated with health care.

\begin{figure}[h]
\centering
\includegraphics[width=1\textwidth]{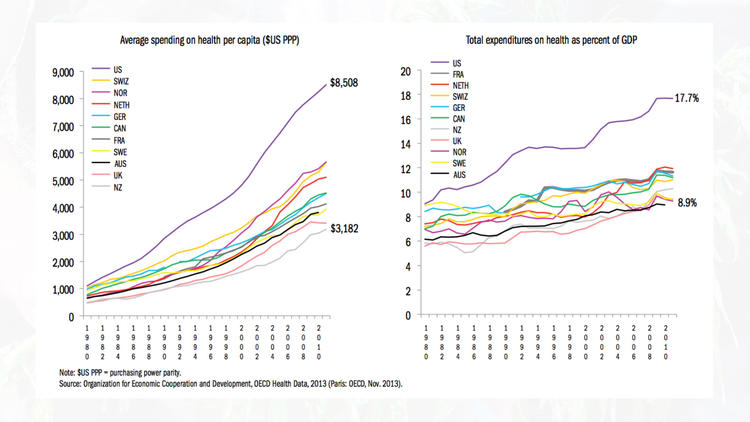}
\caption{Average of spending on health in different countries\cite{LaTimes}}
\end{figure}

	Fig. 1 shows the comparison table of developed countries and the United States spends 8,508 dollars compared to 3,182 dollars in New Zealand. On average, the United States spends 17.7 percent of their expenditure on medical insurance plan while Netherlands spends only 8.9 percent. Although the United States spends a lot of expenditure on health care, the United States has many flaws of the current medical plans. 
    
	     In modern society, people rely on the public health-care system to provide for their medical needs. However, many Americans find that the current health care plan has deteriorated. Due to these crises, many doctors and computer scientists are developing artificial intelligence to find the right treatment for each of the patients in the American health center. I present the future medical system in America with advanced technology adopted in every medical center where patients do not need to worry about not getting appropriate treatment for them. I will explore the Knowledge-Based system that allows doctors to make proper treatments for American patients based on many different medical cases in the medical databases. With “knowledge-based systems”, patients who cannot go to the medical center daily (because they cannot afford to pay the medical bill every time they go or if medical centers are too far away) can use the latest medical treatment can check and monitor their health over the internet, and they can always send their health data directly to their doctors. With the “knowledge-based systems,” I believe that it can be another revolution in our society because previously people had to wait for a long time to meet with their doctors and had to spend a lot of money for treatment. With the “knowledge-based systems,” people do not need to spend a lot of money on spending insurance plans or medical bills anymore. People do not need to overpay for unnecessary care. This paper will explore how a knowledge-based system is built, what algorithms computer scientists use, and how it can change our daily lives over time. 
	A continually large sum of money is needed to grow a medical infrastructure in a country. In the United States, many healthcare companies such as Health Catalyst are adopting new big data technology to provide appropriate treatments for patients \cite{healthcatalyst}. The method of collecting such data and providing results is called a knowledge-based system.

\subsection{Purpose of the Study}
The goal of the study is to conduct a brief overview of the problems with the current health care system and provide solutions to solve each of the problems in the research paper. The solutions involve the area of a knowledge-based system in medical diagnosis founded on the American health care system.
 
\subsection{Significance of the Study}
It is difficult to develop a system that can duplicate the medical diagnosis process of a medical expert that depends mostly on the knowledge obtained through experience. Many attempts have been made to improve the current version of the knowledge-based system, yet the research is going under development. Health Catalyst says, \say{Some academic- or research-focused healthcare institutions are either experimenting with big data or using it in advanced research projects. Those institutions draw upon data scientists, statisticians, graduate students, and the like to wrangle the complexities of big data}. \cite{healthcatalyst}. By experimenting with new health care with recently adopted technologies, patients will be helped with a better solution to cure their diseases and researching medicine to provide a better way to treat undiscovered diseases like middle east respiratory syndrome or other incurable illness like brain cancer.  So how will new technologies like knowledge-based systems with big medical data help to improve the Unites States health care systems? For this purpose, I have undertaken the overview of the medical system in the United States. I am hoping that this will help researchers to create a new prototype of a knowledge-based system and improve the nation's medical infrastructure.

\section{Knowledge-based Systems}
	The knowledge-based system is a method of understanding and developing computational models of human intelligence. \cite{theEssence} ).  In the beginning, the artificial intelligence researchers first developed expert systems. Mostly, the difference of these systems were that \say{the expert system refers to the type of task the system is trying to assist with – replacing or aid a human expert in a complex task} and knowledge-based system \say{refers to the architecture of the system – that it represents knowledge explicitly} \cite{Hayes-Roth:1983:BES:6123}.
 
    	The earlier version of the knowledge-based systems was a basic rule-based expert systems. The earliest version of the rule-based systems thought of this world as a simple assertion and provided how everything can be inserted into a list of rules and how every result has concluded in a certain way, and the experts used the rules to reason about and presented results to add more onto these assertions.

    	The first knowledge-based system called rule-based had some advantages, which were acquisition and maintenance, explanation, and reasoning. By using rule-based, domain experts could define and maintain the rules themselves rather than have programmers build and maintain the standards. By having an explanation, the rule-based system was able to reason about how these systems led to the conclusion and used this information to explain to the users about the conclusion. By doing so, the users were able to utilize the systems to find a diagnosis. The reasoning of the rule-based systems enabled the users to process the knowledge for general purposes. \cite{kbs2}
 
    	As more computer scientists and experts spend time on developing knowledge-based systems, the more knowledge-based systems became increasingly complex regarding how they use some of the reasoning techniques and presenting data to the users becomes more sophisticated. One of the most sophisticated that the systems have become is that the systems became more structured and submitted information by using similar techniques that were occasionally used in object-oriented programming languages such as using a hierarchy of classes and subclasses in a programming language. This method exists in C+ programming language and Java programming language as well as much more. \cite{Hayes-Roth:1983:BES:6123}
 
    	 The most advanced achievement that the scientists have accomplished with the knowledge-based systems is that the scientists were able to adopt the technologies for the development of systems that use the internet. The scientists thought that the internet often has to deal with complex and unstructured data that cannot be utilized for storing information. The technologies that were used in the knowledge-based systems have been adopted to classify which data are reliable and which are not. There are several companies that created new technologies based on the knowledge-based systems such as IBM's Watson and Amazon's Alexa.

\subsection{Basic Components of Knowledge Based System}
A knowledge-based system consists of multiple processes that make the system flow systematically and consistently. A knowledge-based system is software in a computer that collects data and maintains data for presentation purposes. \cite{theEssence}

\subsubsection{Knowledge Base}
The knowledge-base is the main component of the entire knowledge-based system, and this is a place where the domain specific knowledge is captured from the human experts through the knowledge acquisition module. This system consists of many features to represent the knowledge such as frames, knowledge rules, production rules, and semantic rules. This is the center of the knowledge-based system, and it helps store all the important information. \cite{theEssence}

\subsubsection{Inference Engine}
The inference engine is used to perform the task of matching previously collected data to recently added data based on the input given by the users. The most important feature of the engine is to trace its way back through the database of already collected data and then arrive at a conclusion. The benefits of collecting various data are that more knowledge can be included and make the systems more knowledgeable and sophisticated. There are two ways of approaching this procedure: forward chaining and backward chaining. The inference mechanism consists of many different ways of data collecting ranging from IF-THEN statements to case-based reasoning.
 
Two primary methods are used in knowledge-based systems:
 
Forward chaining is one way of the two most important methods of reasoning when using the inference methodology in an IF-THEN module, and it logically follows the rules that an expert made in a system and reaches a conclusion with quantifiable reasons. It is one of the successful reasoning strategies used in the expert system, business, and other production rule systems.
 
Backward chaining is another popular method of reasoning when using the inference methodology in an IF-THEN module, and this method is working backwards and presents how the result have been concluded. \cite{Machine_learning}

\subsubsection{Knowledge Acquisition}
Knowledge acquisition is a subsystem that helps computer or medical experts to build a knowledge base for the entire system. The experts utilize protocol analysis, observation interview techniques for knowledge acquisition. This subsystem is a crucial component of the knowledge-based system. The subsystem collects knowledge that is needed to solve problems and build the knowledge base so that more data can be stored in the system. \cite{review_knowledge}

\subsubsection{Explanation Facility}
An explanation facility is a subsystem that explains the data of the system's actions. In this subsystem, the user can understand how the knowledge-based system concluded its results and how those results were arrived at to justifying the need for extra data to the system. The users may ask basic questions to the system for a more detailed explanation of the conclusion. \cite{review_knowledge}

\subsubsection{User Interface}
The system contains a user interface for communication with the user. The user interface provides some of the important features such as menus, graphical user interface for friendly and convenient access to the database. It also provides converting the rules that were initially programmed by the programmers or experts or newly collected data and other rules from its internal representation to the user for user's interpretation of the data in the system. \cite{review_knowledge}
 
\section{The Current Issues with the American Medical Systems}

\begin{figure}[h]
\centering
\includegraphics[width=1\textwidth]{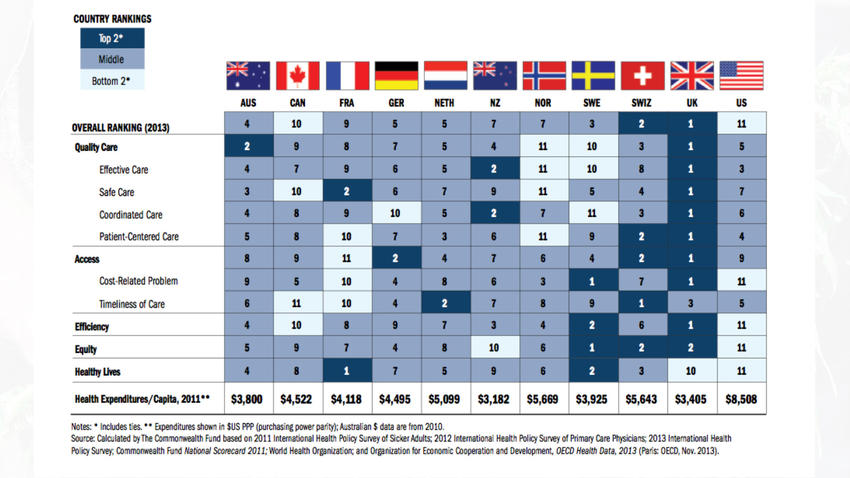}
\caption{Health Care Comparison Chart: The United States has a low quality medical care compared to any other developed countries.\cite{LaTimes}}
\end{figure} 

Even though the United States has a good quality of health care system such as quick response 911 emergency calls and the road is wide enough to transport patients to the closest medical centers. Also many researchers have a better research environment than any other countries so that the researchers or the medical faculties can invest time and money on medical improvement without any limitation. Beyond this, there are many skilled doctors who have good medical experience and medical education backgrounds who are willing to support many patients and provide appropriate treatments. However, in the New York Times, it states \say{That may be true at many top medical centers. However, the disturbing truth is that this country lags well behind other advanced nations in delivering timely and effective care} \cite{world's_best_medical}. The nation has the lowest rate of providing medical care at an affordable price, and it certainly took the lowest rank of the satisfaction of the patients even though the United States has one of the best medical reputations. The United States must enhance its medical services for the wellbeing of its citizens by providing easy access to medical centers and lowering the price of insurance plans so that many American people can afford to pay them. Moreover, what patients get from their medical centers or their doctors may not receive any benefit. It is important for a country to maintain a good health system for its people so that people will have the benefits of medical care whenever they need it. \cite{medicalproblems}

	In the United States, many people have insurance plans. People look at health care insurance plans first to figure out what specific benefits they can get from signing up. One of the U.S.'s healthcare insurance plans called \say{Patient Protection and Affordable Care Act, also known as Obamacare,} \cite{obamacare} has gained critics over the years. According to the definition of the Obamacare, its \say{main focus is on providing more Americans with access to affordable health insurance, improving the quality of health care and health insurance, regulating the health insurance industry, and reducing healthcare spending in the US.} The main goal of the Obamacare was to provide the best-in-class healthcare to every American citizen at an affordable price. However, the new medical plan had many problems that could not be avoided. 
 
\subsection{Problems with Obamacare}
The main idea behind Obamacare was making medical care affordable and allowing many Americans to take benefits of the medical systems of the United States. However, the idea seems not to be that practical in a real-life situation. Obamacare makes people pay extra taxes to the government if they do not have a medical insurance plan. People who did not want to sign up for this medical act had to pay an unnecessary fine to the government, which is unfair to those who are young and healthy, or to those that were not able to get health care for the year because of the medical deadlines for insurance plans. This shows that the United States government has forcefully made their citizens sign up for the medical act and they are just looking to make more money off people. So, what was wrong with the Obamacare?

There are few major issues with Obamacare and these problems are occurring in many places. 

{\bf{A}}. Higher wait times in emergency rooms. Since there are not many rooms for emergency and there is a lack of some doctors to take care of every patient in the nation, patients had to wait for a long time to even get into emergency rooms. This will increase the danger of infection to the patients and it is important to keep patients healthy. 
 
{\bf{B}}. Still, many American people do not have many insurance options, and even worse they do not have many statistics or resources to compare insurance plans. The Forbes used an analogy to illustrate how serious this problem is. \say{We have far more information available to us to compare and select a new car than we do to choose where to go for lifesaving health care}. \cite{Forbes} If U.S citizens have more resources to see more options with their insurance plans, then they could save much money for other important expenses. If there are more resources regarding health care, then the American people would get more benefits. 
 
{\bf{C}}. In the policies of Obamacare, more spending means better medical treatments, which is why the prices of premiums are extremely high. With a common Obamacare plan, the price is already significantly high. However, the premium service has more coverage of health care systems in the nation, and the price was way too high that many American people could not take advantage of this health care plan.

\subsection{Charging People when They Do Not Have Health Care Plans}
Forcing people to spend on unnecessary bills is not good because it is one of the ways to take away the freedom of what people choose to do with their money. The New York Times reports that \say{Balance billing is a controversial and sometimes illegal practice that takes place when doctors and other health-care providers receive a discounted payment from the insurance company — an amount less than the fee they want to be paid} Providing adequate medical care is what every patient is looking for; however, if the new policy fails to do that and keeps providing unnecessary coverage with high cost, then it will make more controversial issues. Here are three basic controversies that the United States faced after the Obamacare.

{\bf{A)}} Obamacare helps to provide free health care, but this did not happen because the United States government forced people to sign up for the new insurance plan even though people already have one and this made many poor people spend additional money in the form of taxes to the government. This is not a practical way of providing adequate medical care.

{\bf{B)}} The bills that were supposed to be free were still paid by the taxpayers, so how can this be free health care in the first place? Obamacare Facts states that \say{Obamacare is free for some via Medicaid expansion, offers cost assistance to others via the marketplace, and cost those without cost assistance money,} but this is strictly restricted to \say{if you or your family qualifies for Medicaid / CHIP}. This is a sign in sheet that determines whether you are qualified to get free medical care. However, there must be another way for making free to non-free. Obamacare managed to solve that by making taxpayers pay for poor people who could not pay for medical care. So, should the government spend money on something that the government forced us to sign up and pay for poor people? \cite{obamacare}
 \cite{obamacare}

{\bf{C)}} Medicaid has already set aside money to pay these unpaid bills, coming from the taxpayer's dollar. Also, Obamacare Facts states that \say{Obamacare isn't free if you are a single low-income adult and your state didn't expand Medicaid, in this case, local charities may be your only option outside a public hospital in an emergency}, \cite{obamacare} so if you are living alone and cannot afford to pay for an insurance plan, then there is no possible way that a doctor comes and assists you. This is an impractical way of helping people who deserve getting the right treatment from the nation.

Obamacare might be a good solution to make the United States’ medical care to be more affordable, but it has many idealistic policies that do not suit every situation just like how they were stated above. To find a better solution, the government should consider a new policy to improve on the current Obamacare and allow more people to get appropriate treatments. So how can it be done?

\section{Solutions}
The United States of America has the best medical system in the world. The nation also has advanced computer technologies that let people stay connected, and this allowed patients to self-diagnose themselves easily by searching on any web engine. However, still, many Americans need to see doctors for a more detailed explanation of their health issues. This paper suggests a solution to shorten the amount of time a patient takes to visit his or her doctor and get an appropriate treatment is simple.
        	
People rely on advanced technologies such as smartphones, tablets, and personal computers every day. Having such advanced technologies with us all the time, we can use those devices to access the artificial intelligence built by science companies like IBM and Amazon. By providing applications on smartphones that every user has and have them input their health data accurately and then send those data to their doctors will allow doctors to monitor their patients' health information regularly. By doing so, patients do not need to worry about keeping in touch with their doctors, and they do not need to worry about waiting for a long period to see their doctors.
        	
With the knowledge-based systems, doctors can fetch data automatically from their patients and manage the data. Incorporating the machine learning techniques like the knowledge-based systems and Bayesian algorithm and Weighted-Majority Algorithm that help the artificial intelligence to safely collect data, present correct results for each of the patients, and even predict symptoms correctly. We can conclude that medical artificial intelligence is the future of American healthcare.

\subsection{To Build Medical Knowledge in Knowledge Based System}
In health care industries, computers are used in a variety of ways. Computers are used to keep patient's health information, details of diseases, details of symptoms, which treatments are given, room reservations, and billing. Moreover, computers are also used in hospital management for registration, appointment with doctors, scheduling, day care units, operating room settings, laboratories, radiology, pharmacy, inventory management system, hospital human resources, revenues, and expenditures of hospitals. Using Artificial Intelligence (AI) techniques, computers can provide the adequate environment for patients and then also provide predicted medical operation procedure for specific patients. With knowledge-based systems built into computers, the systems can even make other alternative medical solutions for patients and provide the percentage of having successful medical operation using Bayesian theorems. 

\subsection{New Decision-Making Strategies}
To prevent a number of mistakes made by doctors when they are operating on patients, doctors proceed with physical examinations on their patients. This procedure provides enough information about the patients to the doctors before the doctors proceed with the surgery. Darzi at the Guardian states that \say{thousands of NHS patients die or suffer serious injuries or infections each year as a result of medical errors} \cite{theguardians}. It is important to reimagine how decision-making strategies are supposed to work. The agents of medical centers who know how to use knowledge-based systems should have a new framework. This way, when doctors are operating, experts can be next to the doctors and monitor the knowledge-based systems to immediately inform the doctors if any problems occur during the surgery.

To collect accurate data from the patients and perform appropriate medical surgery, medical experts should incorporate a new algorithm for every knowledge-based systems in a medical center.  

Here is an algorithm:
\begin{enumerate}
\item Create patient and physical agents
\item Create patient-specific medical data procedure
\item Then, for each timepoint
\item Calculate current outcome delta, physician agent filters evidence
\item Determine optimal current action via medical data procedure search tree 
\item Perform action and update belief states
\item If action is not treat then initiate alternate procedure.
\end{enumerate}
\cite{elsevier_medical}
First, experts should create a patient record including which physical examinations that the patients have gone through. This data will be collected and stored in the knowledge-based systems, and then this will create a patient-specific medical data procedure. Also, from the data collected, the knowledge-based systems can calculate current outcome Delta, which demonstrates an estimated medical surgery procedures, the outcomes of the corrective surgery and the probability of having a successful medical operation.  

From the probability that the systems have concluded, the systems will find evidence that the medical operation will be successful. Also, the systems will provide a conclusion to the doctors and which examination or medical operation would work on the specific patients. This is a revolution of making the medical decision and incorporating medical knowledge-based systems will reduce the number of mistakes that doctors make. 

To illustrate how this process interactive decision-making strategy would work in real life situation, here is a simple rule base:\\

RULE 1:\\
IF the patient has sore throat \\
AND the patient has supprative inflammation of tonsil gland \\ 
THEN the patient is tonsillitis disease \\

RULE 2:\\
IF the patient's temperature is $>$ 37$^{\circ}$ \\
THEN the patient has a fever \\
\cite{diagnosis_medical}

\subsection{New Decision-making Environment}
Creating a new decision-making environment is important for both medical staff and computer experts in a medical center because this will reduce the percentage of making medical operation errors. To reduce the number of medical errors, we need to have a new medical surgery environment with a knowledge-based system in every operating room so that doctors can monitor the current surgery status in real time. The new decision-making environment should contain several requirements to make optimal decisions on every patient without any error because causing another problem may make patients not feel comfortable with the new technology. The knowledge-based systems can fix the current decision-making procedure if doctors can provide enough information to the system. 

\begin{verbatim}
s = state (the actual status of the patients, 
    may contain some information that are not 
    directly observable)
o = observation (patient's observed status/outcome, can be missing)
a = action (treatment option, not treat)
c = treatment costs
CPUC = utilities/rewards
\end{verbatim}
\cite{elsevier_medical}
This will create a network based on the observation of the patients and the network will be stored in the knowledge-based systems as data collecting procedure. This data will be stored safely and doctors can access it remotely. 
's' stands for the currently observed status. If a disease of the specific patient is observed, then the 's' will be assigned with the name of the disease. 'o' is an observation of the patient's status after the physical examination. ‘a’ stands for the treatment option. It lists the number of physical treatment options so that the knowledge-based systems can find which solution would be appropriate for the specific patient. The ‘c’ stands for the cost of treatment. The optimal solution for the cost of treatment would be the outcome of that physical examination or the treatment would cure the patients or not. CPUC is a reward or utilities that how successful the medical operation was. 

Moreover, this information will be accessible to both doctors and patients remotely so patients do not need to spend a lot of time to get their health checked. The patients do not need to make an appointment with their doctors either.  

\begin{figure}[h]
\centering
\includegraphics[width=1\textwidth]{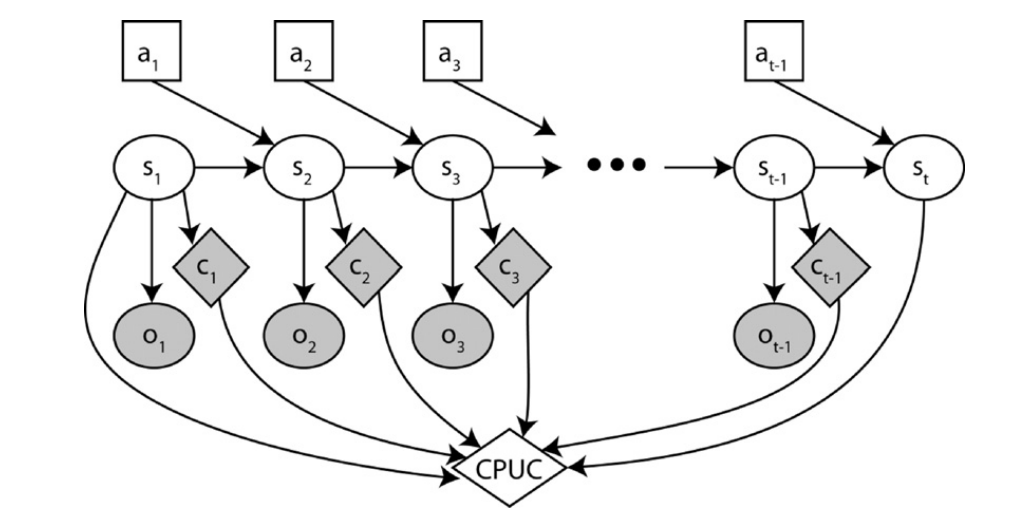}
\caption{A Network of A New Decision-Making Environment\cite{elsevier_medical}}
\end{figure}
This network will be saved in the knowledge-based systems for collecting data purposes and the network will keep growing as more data flow into the knowledge base.

\subsection{New Transformation of Health Care Plans}
Without making changes to the current health care plan, many Americans will not be satisfied. To allow more people to be satisfied with their new health care plan, the United States government should collaborate with medical experts and physicians to accommodate many American people to get diagnosis easily with a lower price for their insurance plans. With a new transformation of health care plans, the government should allow their people to use knowledge-based systems either for free or make the system more affordable and easily accessible through the internet browser or any other form of installation procedures.

With the potential for Obamacare to be repealed, a new system needs to be put in place for the millions who will go without health care if it is repealed. We still need more benefits in the health care system, and we need to integrate the knowledge-based system with the new health care plan.

A. Knowledge-based systems would help with this by alleviating the need for people to have Obamacare in the first place. If people can have access to their doctors remotely wherever they are, then people would not spend a lot of money to transport themselves to a medical center. If people want to have a long summary of their health symptoms, then knowledge-based systems can provide it using its collected data and even suggest how to prevent such symptom from occurring again.

B. A knowledge-based system would transfer data to the doctor about the patient's systems and provide a diagnosis online. Direct contact with the doctors is hard to do at the moment because patients need to wait for a long line outside of doctor's office and many patients find it hard to make an appointment with doctors due to the lack of medical experts in some places in the United States. CNN claims that \say{there aren't nearly enough of us to care for the U.S. population. By some estimates, the country is already short of tens of thousands of doctors, a problem that will only get worse as the demand for care increases with our aging population. That could mean longer wait times for you when you need to make an appointment}. \cite{CNN} With knowledge-based systems, doctors can monitor many patients. For example, one doctor can accommodate a hundred patients at the same time because, if medical experts integrate knowledge-based systems, then the systems will have built-in medical artificial intelligence(AI) and \say{an AI simulation framework can approximate optimal decisions even in complex and uncertain environments}.\cite{healthcatalyst} This means that knowledge-based systems can go through its database and provide appropriate solutions to patients and systems can think like doctors.

C. The only other issue with this is being able to prescribe medications through this system for people to pick up. However, with knowledge-based systems, creating prescriptions will be easy. People can go to their knowledge-based systems to find out which medicine they should purchase from the pharmacy and if we can integrate an express delivery system from UPS or FedEx, then patients can receive their medicine in two business days or less.

\subsection{Make Healthcare More Affordable}
To make health care more affordable, it needs to be easier to access for all classes of Americans.

A. Poorer Americans often cannot afford to go to the doctor, so when they do need to go, their bills often go unpaid. With knowledge-based systems, however, this problem can be solved easily. Poorer Americans do not need to pay their health insurance every moment. Many Americans signed up for healthcare insurance plans just in case they find any health symptoms in their body. Most of these are unnecessary. Forbes explains, \say{Overuse and unnecessary care accounts for anywhere from one-third to one-half of all health care costs, which equal hundreds of billions of dollars, in addition to the half-a-trillion per year experts attribute to lost productivity and disability.} \cite{Forbes}  With knowledge-based systems on people's computers or smartphones, people can browse through the list of symptoms and treatments for those symptoms easily, and they can even navigate the list of medicine that they can purchase from any pharmacy. In addition, people can send direct messages to their doctors through the system.

B. Obamacare helped to alleviate this, by providing free healthcare, but it has many pitfalls, such as making it extremely expensive for small business owners to provide insurance for their employees. To make everyone sign up for health care, the price should be lower. With knowledge-based systems, small business executives can install the software on their currently existing computers and monitor their employees' health easily. Moreover, knowledge-based systems can send employees' health records to doctors remotely, so if there is any health problem discovered in an employee, then small business executives can be notified easily using any smartphone or any computer.
 
C. A combined solution with Obamacare would help to get people insurance so they can go to the doctor, and so that it does not burden the states with debt. Moreover, using information technology like the knowledge-based systems will make patients receive better health care. Surprisingly, many American medical centers are still providing old methods to treat their patients. The Los Angeles Times explains that \say{American primary care doctors lag years behind doctors in other advanced nations in adopting electronic medical records or prescribing medications electronically. This makes it harder to coordinate care, spot errors and adhere to standard clinical guidelines} \cite{LaTimes}. To make the health care systems better, many doctors and medical experts should consider integrating new technologies with their health care systems.

\section{Conclusion}
As this paper went through the research papers from the ACM and USA Today and other scientific papers, it was able to determine some solutions to help Americans. Furthermore, it explained why many computer scientists should consider developing the medical artificial intelligence. It is unfortunate for people who get the wrong treatment due to wrong decisions made by the doctors. Doctors are human, but their mistakes can cost people's lives. To prevent these mistakes, we need to install the knowledge-based systems in every medical center. This way, doctors can remotely access patients' records easily by using on any computer or portable device like phones or tablet PCs to see and analyze every patient in the medical center. By doing so, patients will get proper treatments because knowledge-based systems have an enormous amount of data about each patient and different medical cases so that doctors do not need to worry about making any medical mistakes. However, we should consider taking the health care system seriously and most importantly, the cost of installation. Since every medical center has computational devices and physical treatment devices already, installing knowledge-based system would not be that hard. Also, many knowledge-based systems are software clients, so medical experts can download and install them on their currently existing computers easily. As the Forbes and the ACM claimed, if they could work collaboratively with other governments from different countries and other technology companies in Silicon Valley, then there will be hope for those American patients. This paper focused on the solutions, which will change the way many American's approach brand new medical care and get treatments to cure illnesses they are suffering from. These solutions can make significant changes to the medical care system in America, and the new transformation on health care plans will bring better living conditions to all Americans.

\section{Acknowledgement}
	The author would like to thank Professor Phokion G. Kolaitis, and  Teaching Assistants Kavya Jha, and Edward Rice for recommending and providing knowledge.
%
%

\clearpage
\bibliography{cite}{}

\begin{thebibliography}{10}

\bibitem{healthcatalyst}
Doug Adamson.
\newblock Big data in healthcare made simple: Where it stands today and where
  it’s going.
\newblock
  \url{https://www.healthcatalyst.com/big-data-in-healthcare-made-simple},
  Accessed on February 18th 2017.

\bibitem{Forbes}
Leah Binder.
\newblock The five biggest problems in health care today.
\newblock
  \url{https://www.forbes.com/sites/leahbinder/2013/02/21/the-five-biggest-problems-in-health-care-today/#49884f234587},
  Submitted on 10 Jan 2013.

\bibitem{elsevier_medical}
Kris~Hauserb Casey C.~Bennetta.
\newblock Artificial intelligence framework for simulating clinical
  decision-making: A markov decision process approach.
\newblock
  \url{http://www.sciencedirect.com/science/article/pii/S0933365712001510},
  Available online 31 December 2012.

\bibitem{theEssence}
Alison. Cawsey.
\newblock {\em The Essence of Artificial Intelligence.}
\newblock Harlow: Prentice-Hall, Boston, MA, USA, 1997.

\bibitem{theguardians}
Ara Darzi.
\newblock Doctors make mistakes. the best medicine is for them to admit it.
\newblock
  \url{https://www.theguardian.com/commentisfree/2014/nov/26/doctors-transparency-mistakes-best-medicine-admit-heroic-house},
  Wednesday 26 November 2014.

\bibitem{world's_best_medical}
The New York~Times Editorial.
\newblock World’s best medical care?
\newblock \url{http://www.nytimes.com/2007/08/12/opinion/12sun1.html}, AUG. 12,
  2007.

\bibitem{CNN}
Dr.~Sanjay Gupta.
\newblock America's 9 biggest health issues.
\newblock
  \url{http://www.cnn.com/2015/01/02/opinion/gupta-health-challenges-2015/},
  Mon January 5, 2015.

\bibitem{Hayes-Roth:1983:BES:6123}
Frederick Hayes-Roth, Donald~A. Waterman, and Douglas~B. Lenat.
\newblock {\em Building Expert Systems}.
\newblock Addison-Wesley Longman Publishing Co., Inc., Boston, MA, USA, 1983.

\bibitem{LaTimes}
Michael Hiltzik.
\newblock The u.s. healthcare system: worst in the developed world.
\newblock
  \url{http://www.latimes.com/business/hiltzik/la-fi-mh-the-us-healthcare-system-20140617-column.html},
  June 17th 2014.

\bibitem{medicalproblems}
Dr. Mercola.
\newblock Top ten ways the american health care system fails.
\newblock \url{https://arxiv.org/abs/1403.0835}, 15 March 2014.

\bibitem{kbs2}
William Mettrey.
\newblock An assessment of tools for building large knowledge-based systems.
\newblock
  \url{http://www.aaai.org/ojs/index.php/aimagazine/article/viewArticle/625},
  Accessed February 23, 2017.

\bibitem{Machine_learning}
Tom~M. Mitchell.
\newblock {\em Machine Learning}.
\newblock The McGraw-Hill Companies, Inc., Boston, MA, USA, 1997.

\bibitem{obamacare}
ObamacareFacts.
\newblock What is obamacare | what is the affordable care act?
\newblock \url{http://obamacarefacts.com/whatis-obamacare/}, Accessed on
  Wednesday 15 March 2017.

\bibitem{review_knowledge}
Mrs. Gulavani~S. S. and Kulkarni R.V.
\newblock A review of knowledge based systems in medical diagnosis.
\newblock
  \url{https://www.researchgate.net/publication/228388965_A_REVIEW_OF_KNOWLEDGE_BASED_SYSTEMS_IN_MEDICAL_DIAGNOSIS},
  July-December 2009.

\bibitem{diagnosis_medical}
Watcharachai Wiriyasuttiwong and Walita Narkbuakaew.
\newblock Medical knowledge-based system for diagnosis from symptoms and signs.
\newblock \url{http://www.ijabme.org/images/stories/ijabme/2009/j092009.pdf},
  Published in 2009.

\end{thebibliography}
\bibliographystyle{plain}

\end{document}